\documentstyle [12pt]
{article}
\begin{document}

\hyphenation {never-theless}
\hyphenation {direct-ly}
\hyphenation {ac-ting}
\hyphenation {desing-ularisation}

\title{More about discrete symmetries in compactified string theories}
\author{Christoph
M.A. Scheich\thanks{Supported by the
European Community, Science Program},
\\Department of Physics,\\
Theoretical Physics,\\
University of Oxford,\\
1 Keble Road,\\
Oxford OX1 3NP}
\maketitle
\begin{abstract}
\noindent
We discuss discrete symmetries in several string compactification
schemes. The same constraints on the light spectra as for Gepner
models \cite{rosss} are found in various cases for non-$R$
symmetries. The analogous
constraints for $R$ symmetries are also established. Therefore it seems
natural to conjecture that they always apply.
\end{abstract}
\maketitle

In a previous paper we found constraints on the light
spectra of string compactifications using $N=2$ minimal superconformal
models (Gepner models). These constraints are very similar to those that
were
found for discrete gauge symmetries in field theories \cite{ibanez}.
Here we want to discuss the results of \cite{rosss} in more detail,
especially in comparison to the field theory result.

Furthermore we find the same constraints fulfilled for phenomenologically
interesting models in other compactification schemes as for Gepner
models. In particular we will discuss implications for Calabi-Yau,
orbifold and fermionic string constructions.
We also extend our consideration of non-$R$ symmetries to $R$
symmetries along the lines of \cite{ibanezneu} and find the
analogous constraints fulfilled.

To start with we discuss the discrete symmetries in Gepner
models in some more detail.

The Gepner models in $D=4$ space-time dimensions are heterotic
string models in which compactification is achieved by tensoring
$N=2$ minimal superconformal models to form an internal sector
of the theory in such a way that  conformal anomaly
cancellation is achieved and a $N=1$ space-time supersymmetric,
modular invariant theory with correct spin-statistics is
generated \cite{gepner}. These models are characterized by their
level $k_i$ and their type according to the $ADE$-classification
scheme  \cite{ade}.
They have conformal anomaly
\begin{equation}
c_i\ =\ \frac{3k_i}{k_i+2}\ \ .
\end{equation}
The states of the theory are formed from tensor products of the
primary fields
$\Phi^{l;\overline{l}}_{q,s;\overline{q},\overline{s}}$
associated with each factor. The conformal dimensions and $U(1)$-
charges of these fields are given by
\begin{equation}
h\ =\ \frac{l(l+2)-q^2}{4(k+2)}\ +\ \frac{s^2}{8}\ \ ,
\end{equation}
\begin{equation}
Q\ =\ -\frac{q}{k+2}\ +\ \frac{s}{2}\ \ .
\end{equation}
The heterotic string theory then possesses in general a gauge
symmetry $G\otimes E_8'\otimes U(1)^{r-1}$, where $G$ denotes
an embedding only into $E_6$, which we assume for the moment.

For the left moving part of the string the $U(1)$ gauge group
of each tensor factor $i$ contains a discrete subgroup of
non-$R$ symmetries $Z_{2(k_i+2)}$ if $k_i+2$ is odd or
$Z_{k_i+2}$ if $k_i+2$ is even, with charges
\begin{eqnarray}
\sum_{J=1}^n \bar{q'}_i^{J} \equiv
\left\{ \begin{array}{lll}
\sum_{J=1}^n
[-2\bar{q}_i^J + (k_i+2)\bar{s}_i^J] = 0 & mod\ 2(k_i+2) &
\rm for\it\ k_i+\rm 2 \ odd\it
\\
\sum_{J=1}^n [-\bar{q}_i^J + \frac{k_i+2}{2}\bar{s}_i^J] = 0
                                         & mod\ k_i+2 &
\rm for\it\ k_i+\rm 2 \ even\it\ \ (\forall i )\ ,
\end{array} \right.
\label{neulad}
\end{eqnarray}
where $n$ is the number of fields appearing in the correlation
function \cite{andy}. Usually one splits them apart
by writing as a $Z_{k_i+2}\otimes Z_2$ symmetry
\begin{eqnarray}
\sum_{J=1}^{n}\bar{q}_{i}^{J} =0 & mod\ k_i+2 &
(\forall i)\ ,
\label{s1}\\
\sum_{J=1}^n \bar{s}_i^J =0 & mod\ 2 &
(\forall i)\ .
\label{s11}
\end{eqnarray}
The right moving part of the string has a
$\tilde{Z}_{k_i+2}\otimes \tilde{Z}_2$ symmetry which is generically
a $R$ symmetry
\begin{eqnarray}
\sum_{J=1}^n q_i^J + 2 = 0 & mod\ k_i+2 &
(\forall i)\ ,
\label{s2}
\\
\sum_{J=1}^n s_i^J + 2\Sigma_{J=4}^n d_i^J = 0 & mod\ 2 &
(\forall i)
\label{s22}
\end{eqnarray}
and follows from the selection rules for non-vanishing correlation
functions involving $n$ fields which come from the parafermionic
nature of the primary fields \cite{kopp}.
Here $d_i^J$ describes the Vertex operators in the $(0)$ picture.
We adopt the convention that the quantum numbers $q$, $s$ refer
to the fields in the $(-1)$ picture and the $\bar{q}$, $\bar{s}$
refer to the
appropriate choice of representation leading to a gauge invariant
correlation function. In the case of $A$, $D_{odd}$, $E_6$ modular
invariants the diagonal sum and difference of eqs~(\ref{s1})
and
(\ref{s2}) is a $Z_{N_i}\otimes {Z'}_{N_i}$, $N_i=2(k_i+2)$ symmetry
and that for (\ref{s11}) and (\ref{s22}) a $Z_4\otimes Z'_4$ symmetry
\cite{andy}.
For $D_{even}$, $E_7$, $E_8$ modular invariants the
diagonal sum and differences constitute only a $Z_{N_i}\otimes
Z'_{N_i}$, $N_i=k_i+2$ and a $Z_2\otimes Z'_2$ symmetry.

But the Gepner construction realizes only a subgroup nontrivially.
As explained in \cite{wir} and below one gets a non-$R$ symmetry
\begin{equation}
G_{\rm non\it -R} = \prod_{i=1}^r Z_{N_i/2}/Z_{\tilde{N}/2} (\times S)\ ,
\end{equation}
where $\tilde{N}$ denotes the smallest common multiple of all $N_i$ and
we wrote the permutational symmetries in brackets (which will not be
discussed any further). The $R$ symmetry
\begin{equation}
G_R = Z_2
\end{equation}
describes if states are in the Ramond or Neveu-Schwarz sector. But
we should comment that the nature of the correlation functions is
such that there are many further \it stringy\rm\ symmetries.
This implies that
fields have only relative quantum numbers and not absolute ones.
Therefore symmetries
exist that can not be associated to symmetries in
a low energy field theory, but still restrict
couplings\footnote{For a similar discussion see \cite{select}.}.
\hfill \break
The diagonal sum of charges
divided by 2 are
the socalled Gepner charges defined for the $\underline{10}_{scalar}$
representation. This
is the symmetry that is found in Calabi-Yau constructions
associated to superconformal theories.

In the following let us start with a study of the $\bar{q'}$,
$q$, $\bar{q}$,
$s$, $\bar{s}$ charges and then apply the result to the linear
combinations (of left and right moving charges).
\hfill \break
The $\bar{q'}$ charges fulfill the relations $i$ to $iv$ of
\cite{ibanez} for spontaneously broken Abelian gauge symmetries
immediately, since the $Z_{2(k_i+2)}$ or $Z_{k_i+2}$
symmetry is a subgroup of the $U(1)$s, which are anomaly free.
Furthermore the term due to Majorana masses is lacking.
This is obviously independent from the specific basis choosen
for the charges.
\hfill \break
To deduce the relations for the symmetries of the $\bar{q}$
charges we may choose an arbitrary basis and use again
the connection of these symmetries with the $U(1)^{r}$ gauge
symmetries of the Gepner construction. The latter are all anomaly
free and so the ($M$) massless states of the theory satisfy
the
mixed $U(1)$ gravitational anomaly cancellation conditions
implying immediately

i) The mixed $Z_N$\it -G-G\rm\ gravitational
anomaly conditions
\begin{equation}
\sum_{J=1}^M \bar{q}_i^J = pN + \eta q_q\frac{N}{2}\ \ p,q\in Z,
\label{nur}
\end{equation}
where $\eta_q$=1,0 for $N$ even, odd. Following alone
from the fact that $\sum_{J=1}^M\bar{q}_i$ has to be an integer.
\hfill \break
Using the same arguments it is now straightforward to derive the
remaining anomaly cancellation conditions for the discrete
symmetries in the same basis as above:

ii) Pure discrete $Z_NZ_MZ_L$ anomaly cancellation condition:
\begin{equation}
\sum_{J=1}^{M}\ q_i^{J}p_i^Jo_i^J\ =\ r'N+s'M+t'L+\eta_Nu'\frac{N}{2}
                                +\eta_Mv'\frac{M}{2}
                                +\eta_Lw'\frac{L}{2}
                      \ \ ,\ \ r',s',t',u',v',w'\ \in\ Z,
\label{eq:u13}
\end{equation}
where $\eta_N$, $\eta_M$, $\eta_L$ = $1,0$ for $N$, $M$, $L$
even, odd and $q_i$, $p_i$, $o_i$ are the discrete charges
($=\bar{q}_i^J$).
The last terms are in general different from the result of \cite{ibanez},
since they do not vanish automatically if one of the discrete groups is
odd. In \cite{ibanez} the reason being no Majorana mass terms for odd
groups, which has no correspondence on the charge level.
\hfill \break
We should stress that in contrary to \cite{ibanez} the size
of the discrete groups are fixed due to the underlying string
theory.

iii) Mixed discrete and gauge anomalies:
\begin{equation}
\sum_{J=1}^M\ T_i^Jq_i^J\ =\
\frac{1}{2} r''N\ +\ u''\eta_N\frac{N}{4}
\ \ , \ \ r'',u''\ \in\ Z\ \ ,
\end{equation}
where $T_i^J(R)$ is the quadratic Casimir corresponding to each given
representation $R$ (the normalization is such that the Casimir for
the $M$-plet of $SU(M)$ is $\frac{1}{2}$).
Also differing slightly from \cite{ibanez}.

iv) Mixed discrete and $U(1)$ anomalies of types
$Z_NU(1)_XU(1)_Y$ and $Z_NZ_MU(1)_X$:
\hfill \break
\begin{eqnarray}
\sum_{J=1}^M\ x_i^Jy_i^Jq_i^J\ =\ r'''N\ +\ u'''\eta_N\frac{N}{2}\ \ ,
\ \ r''',u'''\ \in\ Z\ \ ;    \label{least} \\
\sum_{J=1}^M\ x_i^Jq_i^Jp_i^J\ =\ r^{iv}N+s^{iv}M+\eta_Nu^{iv}\frac{N}{2}
                     +\eta_Mv^{iv}\frac{M}{2}
                \ \ ,\ \ r^{iv},s^{iv},u^{iv},v^{iv}\
\in\ Z\ \ , \label{last}
\end{eqnarray}
where $x_i^J$, $y_i^J$ denote the $U(1)$ charges. Again the last two
terms are in general different from the result in \cite{ibanez}.
\hfill \break
Note that in the case of discrete gauge
symmetries this condition can only be derived once one knows the
normalisation of the underlying $U(1)$ currents i.e. in this case
one must know the transformation properties of the massive as
well as the massless states. For this reason one could not use
this sum rule to constrain low energy models in \cite{ibanez}. In
the present case, however, we have been able to derive the
sum rules eqs.~(\ref{least}) and (\ref{last})
for the light states only because the 4D-
string construction delivers normalised $U(1)$
currents.
\hfill \break
As such $i$ and $iv$ provide stronger constraints on the
possible low energy theories with discrete symmetries descending
from the superstring.

These relations differ
 from the relations of the $\bar{q'}$ charges by the
$\eta$ terms. Let us explore the possibility of getting vanishing
$\eta$ terms to compare further with \cite{ibanez}.
\hfill \break
In the case that the original $E_6(\otimes E'_8)$ gauge symmetry
is not broken by twists one
derives the discrete
anomaly cancellation conditions $i$ to $iv$ without the $\eta$ terms,
because of the multiplicities of the states in the $SO(10)$
representations \cite{rosss}.
In the case that the original $E_6(\otimes E'_8)$ gauge symmetry
is broken by
twists and embeddings (see below for details), $\eta$ terms appear
in general.
\hfill \break
But now the question arises, if it is possible to form linear
combinations of charges such that a maximum number of $\eta$ terms
vanishes (like one also does for anomalous $U(1)$ gauge symmetries
in string theories
and finds that at maximum always one of them
is anomalous, what is associated with the breaking of the $E'_8$
\cite{priv}).
It is
convenient to introduce a basis for the discrete symmetry
generators in such a way that contributions for odd $\bar{s}_i$
vanish. From the absence of the $U(1)$\it -G-G\rm\ anomaly
\begin{equation}
\sum_{J=1}^M [-\{(k_1+2)\bar{q}_i-(k_i+2)\bar{q}_1\} +
\frac{1}{2}(k_i+2)(k_1+2)\{\bar{s}_i-\bar{s}_1\}] = 0.
\label{super}
\end{equation}
Thus we have
\begin{equation}
\sum_{J=1}^M (k_1+2)\bar{q}_i-(k_i+2)\bar{q}_1 = pN_i + p'N_1,
\hspace{1cm}p,p' \in Z,
\end{equation}
where $N_i=(k_i+2)$. For the case $N_i$ and $N_1$ have a common factor
$(k_1+2)\bar{q}_i-(k_i+2)\bar{q}_1$ is a generator of a discrete
symmetry and eq.~(\ref{super}) gives the mixed $Z_N$\it-G-G\rm\ gravitational
anomaly cancellation condition without a \it Majorana term\rm , only
for this common factor. The same applies for the relations
$i$ to $iv$. By taking appropriate combinations
$(k_1+2)\bar{q}_{i}-(k_i+2)\bar{q}_{1}$ of the set
all discrete symmetries in the new basis
may be obtained. For the residual discrete symmetry
which may be chosen to be generated by $q_1$, we
are only able to prove the weaker conditions (\ref{nur}) to
(\ref{last}).
\hfill \break
But after all such a basis guarantees anomaly cancellation only for
common factors, thus being in general insufficient. Nevertheless
for the phenomenologically interesting models being relevant
(see below).

So far we have discussed the anomaly cancellation conditions
associated with the $Z_{N}$ symmetries of the $\bar{q}$
charges. The $Z_2$ symmetry associated to the $\bar{s}$
charges is a subgroup of the gauge group.

The anomaly cancellation condition for the
$\tilde{Z}_{N}$ symmetries associated with the right sector
charges (cf. eq.~(\ref{s2})) follow from these results by using
the Gepner construction of massless states. Let us first discuss the
case of untwisted models. One starts with a
combination of half integer charge $Q_{tot}$ and appropriate
conformal dimension $h$ in the right internal sector and obtains
the ones in the left sector by adding ($n$) multiples of
$\beta_0$, and ($m$) multiples of $\beta_i$ \cite{gepner} in such
a way that the resulting states have again half integer
$\overline{Q}_{tot}$ and appropriate $\overline{h}$. Here
$\beta_0$ is the generator of supersymmetry and acts by adding
1 to each $q_i$ and $s_i$ component. The vector $\beta_i$ only
acts by adding 2 to the $i$th $s_i$ index. Thus
\begin{equation}
q_i^J=\bar{q}_i^J -n\ \ \  ,\ \ \
s_i^J=\bar{s}_i^J -n-2m_i\ \ \ (\forall i).
\label{lr}
\end{equation}
Now if we define a right analogon $q'_i$ of the $\bar{q'}_i$
charge as in (\ref{neulad}) and choose linear combinations
$(v_1,\cdots ,v_r)$ containing $\beta_0$ itself such that
\begin{eqnarray}
\sum_{i=1}^r \frac{k_i}{k_i+2} v_i^mv_i^n = Z & \forall\ m,n
\ \ ,
\end{eqnarray}
then $q'$ and $\bar{q'}$ charges except for the $\beta_0$ combination
differ only by multiples of the appropriate $\tilde{N}$.
Therefore
all the combinations except the
$Z_2$ $R$ symmetry $\beta_0$ fulfill $i$ to $iv$ without
$\eta$ terms\footnote{We should remark that a verification
of the gravitational and gauge anomaly cancellations is
most conveniently performed in such a basis for the Gepner
models.}.

For the $q$ and $\bar{q}$ charges we can not expect to be able
using the same basis and indeed it is convenient to introduce
the following one. One chooses
orthogonal combinations of the vectors $(v_1,\cdots ,v_r)$ that
contain (the $q$ part of) $\beta_0$:
\begin{equation}
\sum_{i=1}^r \frac{v_i^mv_i^n}{k_i+2} = Z\ ,\ \ \forall
\label{eq:basis}
\ m,n.
\end{equation}
So we get charges that commute with $\beta_0$, except the diagonal
sum. Furthermore these combinations mean that
all of them are non-$R$ discrete symmetries, except the diagonal
sum of the $q_i$ which is a  potential $Z_2$ $R$ symmetry (if
not trivial).
Thus in this basis it is clear that the left and right discrete
charges are the same and the anomaly cancellation conditions for
the right sector are the same as for the left for the combinations
commuting with $\beta_0$.

The possible $Z_2$ $R$ symmetry associated to the $\beta_0$
combination of $q$ charges or $s$
charges\footnote{Depending on the number of $N=2$ minimal
superconformal models being even or odd.} have
to fulfill relations different from $i$ to $iv$. So far only the
ones except the pure discrete ($ii$) are available from \cite{ibanezneu}.
We apply these for the ones in the Gepner model.
If the $Z_2$ is associated to $\beta_0$ the charges $q_g$ of gravitinos and
gauginos are the number of the $N=2$ minimal models $r$. For the
$R$ symmetry associated to $s$ charges it depends on
the actual basis. We expect in analogy to \cite{ibanezneu} for $Z_N$
symmetries

i')
\begin{equation}
\sum_{J=1}^Mq_i^J+(dimG-21)q_g\ =\ rN+\eta_N n\frac{N}{2},\ \ n,r\in Z,
\end{equation}
where $dimG$ is the dimension of the complete gauge group, $q_g$
the charge of the gaugino resp. gravitino and $N=2$ in our case.

iii')
\begin{equation}
\sum_{J=1}^M T_i^Jq_i^J+C_2q_g\ = \ \frac{1}{2}r'''N+
s'''\eta_N\frac{N}{4},\ \ r''',s''' \in Z,
\end{equation}
where $C_2$ is the second Casimir invariant
being $M$ in the case of $SU(M)$ and again $N=2$. Here one has very
often vanishing $\eta$ term.
\hfill \break
These sum rules
are different from the ones for non-$R$ symmetries, while $iv'$ is
the same as for non-$R$ symmetries with the $\eta$ terms as above.
In relation to \cite{ibanezneu} we get also the additional term at the
rhs.

In the case of twisted models the Gepner construction is slightly
more complicated. Instead of eq.~(\ref{lr}) we then have
\begin{equation}
q_i^J=\bar{q}_i^J-n-2pt_i\ \ \ ,\ \ \
s_i^J=\bar{s}_i^J-n-2m_i\ \ \ (\forall i)\ ,
\label{twist}
\end{equation}
where $t_i$ is the twistvector \cite{gepner} and $p$ an integer
characterizing the twistsector.
To commute with
supersymmetry ($\beta_0$) it fulfills
\begin{equation}
\ \sum^r_{i=1}\frac{t_i}{k_i+2}\ \in\ Z\ .
\label{susy}
\end{equation}
Furthermore for the allowed states we have the additional projection
\begin{equation}
\tau\bullet (V+\overline{V})\ -\ (\delta_o,2\alpha+p\delta_p)\ \in\
Z\ ,
\label{con}
\end{equation}
where $\tau_i=2t_i$; $\delta_o$, $\delta_p$ characterize the embeddings
and $\alpha$ are the vectors that characterize the representation.
(See e.g. \cite{wir}
for the conventions.)
We simply have to include $t_i$ in the above basis of charges as one of
the combinations.
Then the last term in eq. (\ref{twist}) is always trivial and all
conclusions from the untwisted case apply.
\hfill \break
For only phasetwisted models (trivial embedding), one is
able to show that there are further discrete symmetries $Z_t$
associated
with the twistsectors of a given model which have $p$ as a charge
 (called \it twistnumber\rm\ symmetry in the following) and
that fulfill $i$ to $iv$.
\hfill \break
In the case of nontrivial embeddings $\delta_o$, $\delta_p\neq 0$ also
the second term on the left of eq.~(\ref{con}) is present. The above
result immediately extends - but not for the twistnumbers, where the proof
is not valid anymore.
Nevertheless several examples inspected by us have shown no violation
of $i$ to $iv$ for $Z_t$.

Now we want to address the question of linear combinations of charges.
\hfill \break
$q'_i+\bar{q'}_i$ multiplies all charges by 2 and gives $i$ to $iv$
while $q'_i-\bar{q'}_i$ is trivial.
\hfill \break
Let us take the $Z_{2(k_i+2)}\otimes
{Z}_{2(k_i+2)}$ resp. $Z_{k_i+2}\otimes Z_{k_i+2}$
charges $q_i+\bar{q}_i$ and $q_i-\bar{q}_i$.
In our basis (\ref{eq:basis})
commuting with $\beta_0$, we see immediately that $\bar{q}+q$
 just multiplies the $\bar{q}$ charges by 2 and we get the relations
$i$ to $iv$ for $N'=2N$ symmetries. Therefore it is actually only a $N$
symmetry. This being the reason one is allowed to introduce the
Gepner charges $\frac{q_i+\bar{q_i}}{2}$ fulfilling $i$ to $iv$ as a
$Z_{k_i+2}$  resp. $Z_{(k_i+2)/2}$ symmetry.
The combinations $q-\bar{q}$ in our basis vanish. For the
$Z_2$ $R$ symmetry just redefinitions take place.
\hfill \break
To be specific we consider now two phenomenologically interesting models.
We start with the $(0,1,2,3,4)$ modded $(3_A,3_A,3_A,3_A,3_A)$ model.
Relations $i$ to $iv$ are fulfilled for a $Z_{5}^4$.
The same is true for the twistnumber
$Z_5$ symmetry and the $Z_2$ $R$ symmetry of $s$ charges satisfies
$i'$ to $iv'$.
\hfill \break
For the $(0,6,12,0)$ modding of the $(1_A,16_E,16_E,16_E)$ model
a $Z_3\otimes Z_{18}^3$
fulfills $i$ to $iv$. This is further reduced to a $Z_3\otimes Z_{9}^3$
by the fact that all $q$ and $\bar{q}$ charges are even.
The $Z_2$ $R$ symmetry generated by
$\beta_0$ satisfies the relations $i'$ to $iv'$ trivially.

Generalization to embeddings also into $E'_8$ is straightforward
and one has to find a suitable basis to isolate the anomaly.

So we see that the non-$R$ discrete symmetries generated by the
$q'$ and $\bar{q'}$ charges in Gepner models fulfill the same
conditions as were necessary for the discrete anomaly cancellation
in field theories to stabilize them against quantum gravity corrections.
They are even stronger than those.
The reason for this relations in Gepner models being the additional $U(1)$
currents in the $N=2$ minimal superconformal models.
For the symmetries generated by $q$, $\bar{q}$, $s$, $\bar{s}$ only
relations $i$ to $iv$ apply. For $R$ symmetries $i'$ to $iv'$ are valid.

Since almost every Gepner model is related to a Calabi-Yau
manifold\footnote{For a discussion of this relations see e.g. \cite{asp}.},
we can gain
information also about them. If one moves away from this point in moduli
space by giving VEV's to complex structures and radii, $U(1)$s and discrete
symmetries are broken either on the left and right moving side to $Z_N$'s,
where the VEV's have charges $qN$. Because of the masses created by the
VEV's, the equations $i$ to $iv$ have to be changed including nontrivial
terms due to massive states. Thus the $U(1)$s and discrete symmetries are
simultaneously broken and the remaining symmetries fulfill $i$ to $iv$
also when moving to a Calabi-Yau manifold. Then the same conclusion about
the symmetries usually considered for Calabi-Yau manifolds apply as
above\footnote{We keep in mind that the VEV's for the radii lower
discrete symmetries although this is not obvious from the manifold
formulation.}.
\hfill \break
But there are now much more couplings that may be corrected by
quantum
gravity. Thus the points in moduli space associated with the quantum theories
seem to be favoured concerning the question for the absence of quantum
gravity corrections that may make the model unrealistic. For Calabi-Yau
manifolds to be a starting point of a prediction of low energy theories,
it seems preferable to be similar to an underlying conformal field
theory.

Let us now discuss the discrete symmetries in the case of the
fermionic string \cite{fermionic}. The symmetries of the model
are the various gauge symmetries and the discrete symmetries
given by the vectors $W_i$ that characterize the boundary
conditions of the fermions. \hfill \break
They describe the transformation under
$\sigma_1\rightarrow\sigma_1 +2\pi$ \begin{equation}
\psi^i(\sigma_1+2\pi ,\sigma_2)\ =\ e^{-2\pi i W_i}
\psi^i(\sigma_1,\sigma_2) \ \ ,
\label{eq:transform}
\end{equation}
where $\sigma_1$, $\sigma_2$ are the world-sheet coordinates
(like the twists in the Gepner model). So one has a discrete
$Z_N$  \it twistnumber\rm\ symmetry for each
fermion under which e.g. the correlation functions have to be
invariant \cite{balin}. There is a fermionic charge associated
\begin{equation}
Q_i\ =\ N_i\ +\ W_i\ -\ W_{0i}\ \ ,
\end{equation}
where $N_i$ is the fermionic number operator and
$W_{0i}=\frac{1}{2}$ for all $i$. In the case that there are
$U(1)$s (possibly embedded in bigger gauge groups) these are the
charges under the $U(1)$s. In this case one has  anomaly
cancellation for the gauge symmetries which implies $i$ to $iv$
for the discrete $Z_N$ symmetries which are contained
in them (like for the left moving sector of the Gepner model).
\hfill \break
In the case of a broken $E'_8$
gauge group
one has one anomalous $U(1)$ factor, for which the corresponding
$Z_N$  may also be anomalous.
\hfill \break
For the \it twistnumbers\rm\ associated to the left movers no anomaly
free $U(1)$ symmetries that would lead to $i$ to $iv$ exist.
Let us give the result for the $SU(5)$ flipped model as an
example  \cite{ellis}. The discrete symmetries of the model
fulfilling $i'$ to $iv'$ and $i$ to $iv$
are: a $Z_2$  symmetry stemming from the anomalous $U(1)$ and the
$Z_2$ symmetries  associated with the twist vectors $S$,
$b_1+\zeta$, $b_2+\zeta$, $b_3+\zeta$, $b_4+\zeta$, $b_5+\zeta$,
$b_1+b_2+b_3+b_4+b_5-\zeta$ in the notation of
\cite{ellis}\footnote{This time without a \it Majorana mass\rm\ term for
the non-$R$ symmetries.}. Discrete symmetries associated with the $Z_4$
symmetry $\alpha$ do not. Such a discrete anomaly does not allow
for quantum gravity corrections of the couplings between the
crucial matter fields, since they have zero $\alpha$-twist and
are thus not affected.

Futhermore we want to make a remark on
 the discrete symmetries of $Z_N$ orbifolds.
The authors of \cite{select} have found four selection rules that are
associated to three classes of symmetries:
\hfill \break
1. The space group $S$ selection rule contains the point group $P$
selection rule which is the rotation part of $S$. The discrete symmetry
associated with it is that fields in a twistsector $\Theta^m$,
$m=0,\cdots ,N-1$ are multiplied by a phase $exp(2\pi im/N)$ similar to
(\ref{eq:transform}). In the case of $Z_N$ orbifolds this is a $Z_N$
symmetry.
\hfill \break
2. The conservation of $H$-momenta and the invariance under independent phase
rotations of orbifold coordinates. Together they form discrete $R$-symmetries
that can e.g. be read of from the Vertex operators of the particles: a
vertex operator involving exponentials $exp(i\beta_i H_i)$ and powers of
derivatives $(\partial X)^{a_i}$, $(\overline{\partial}\overline{X})^{b_i}$
represents a particle with charge $\beta_i+a_i-b_i$, where $i$ denotes the
set of independent phase rotations.
\hfill \break
3. The selection rule 4 of \cite{select} corresponds to \it stringy\rm\
zeroes
and cannot be understood as a discrete symmetry of a field theory.
\hfill \break
Since the reasons for fulfilling $i$ to $iv$ seem more involved, we
consider here an example.
Let us just check 1 and 2 for a very prominent case: the
three generation $Z_3$ orbifold (e.g. with gauge group
$SU(3)\otimes SU(2)\otimes U(1)^8\otimes SO(10)'$
\cite{sierra}. In this case
1 is a $Z_3$ and 2 a $Z_3^3$ symmetry. The anomaly cancellations $i$ to
$iv$ are fulfilled for the non-$R$ symmetry,
 since every state of the massless spectrum comes in
3 copies with the same quantum numbers. The analogous constraints for
$R$ symmetries, i.e. $i'$ to $iv'$, are also fulfilled for the same
reason.
\hfill \break
Since the couplings of this model are mainly determined by this discrete
symmetries, the discussion of the three generation $Z_3$ orbifold in
\cite{sierra} remains unaffected.

Finally we can say that we have found that relations $i$ to $iv$
are fulfilled for a specific basis in Gepner models without the
$\eta$ terms. We also verified the same for a phenomenological
interesting orbifold and fermionic string model\footnote{In the last
case at least in the observable sector.}. Relations $i'$ to $iv'$
were also found fulfilled for $R$ symmetries in Gepner models, fermionic
strings and orbifolds.

Low energy theories stemming from string theories are
therefore conjectured to
fulfill relations $i$ to $iv$ and $i'$ to $iv'$.
The reason for fulfilling not the
stronger relations with vanishing $\eta$ terms simply being the fact that it
is not always possible to find the appropriate basis. For example
for a Calabi-Yau manifold only the charges for fulfilling $i$ to
$iv$ in general are known. Furthermore sucessive spontaneous
breaking of symmetries below the compactification scale leads always
to terms as found in \cite{ibanez}. Since relations $i$ to $iv$
contain those and being weaker they have always to be fulfilled.
Finding stronger relations for discrete symmetries than in
the field theory case could mean that these symmetries are
remenants of the symmetries with vanishing $\eta$ terms.

\vspace{1cm}
\noindent
{\bf \Large Acknowledgement} \hspace{0.3cm} I like to thank
G.G. Ross for intensive discussions and L.E. Ibanez for
pointing out a mistake.


\newpage \begin{thebibliography}{99}
\newcommand{\npb}{\mbox{Nucl.Phys.B }}
\newcommand{\pl}{\mbox{Phys.Lett.B}}


\bibitem{rosss} G.G. Ross, C.M.A. Scheich, to appear soon
\bibitem{ibanez} L.E. Ibanez, G.G. Ross, \pl 260 (1991) 291
\bibitem{ross+} L.E. Ibanez, G.G. Ross, \npb 368 (1992) 3
\bibitem{krauss} T. Banks, Santa Cruz preprint SCIPP 89/17 (1989);
                L. Krauss, F. Wilczek, Phys.Rev.Lett. 62 (1989)
1221;                  T. Banks, \npb 323 (1989) 90;
                 L.Krauss, Gen.Rel.Grav. 22 (1990) 50;
                 M. Alford, J. March-Russell, F. Wilczek, \npb
337                  (1990) 695
\bibitem{adler} S.L. Adler, W.A. Bardeen, Phys.Rev. 182 (1969)
1517;                 D.J. Gross, R. Jackiw, Phys.Rev.D 6 (1972)
477
\bibitem{ade} A. Capelli, C. Itzykson, J.G. Zuber,
Commun.Math.Phys. 113               (1987) 1
\bibitem{kopp} V.A. Fateev, A.B. Zamolodchikov, Sov.Phys. JETP
62(2),                August 1985
\bibitem{andy}C.A. L\"utken, G.G. Ross, \pl 214 (1988) 357,
              S.F. Cordes, Y. Kikuchi, Mod. Phys. Lett. A 4
              (1989) 1365
\bibitem{gepner} D. Gepner, \npb 296 (1988)
757; \pl 199 (1987) 380;                  Princeton preprint,
December 1987
\bibitem{priv} L.E. Ibanez, private communication
\bibitem{fermionic} H. Kawai, D.C. Lewellen, S.H.H. Tye, \npb 288
(1987) 1;                     I. Antoniadis, C.P. Bachas, C.
Kounnas, \npb 289 (1987)                     87
\bibitem{wir} J. Fuchs, A. Klemm, C. Scheich, M.G. Schmidt, Ann.
of Phys.               204 (1990) 1
\bibitem{asp} R. Schimmrigk, Phys.Rev.Lett. 70 (1993) 3688;
              P.S. Aspinwall, C.A. L\"utken, OUTP-90-12P
\bibitem{ibanezneu} L.E. Ibanez, CERN-TH.6662/92, September 1992
\bibitem{balin} D. Bailin, D.C. Dunbar, A. Love, \pl 219
(1989) 76
\bibitem{ellis} I. Antoniadis, J. Ellis, J.S. Hagelin, D.V.
Nanopoulos,         \pl 231 (1989) 65; J. Rizos, K. Tamvakis, \pl
251 (1990) 369
\bibitem{select} A. Font, L.E. Ibanez, H.-P. Nilles, F. Quevedo,
\npb 307 (1988) 109 and \pl 213 (1988) 274
\bibitem{sierra} L.E. Ibanez, J.E. Kim, H.-P. Nilles, F. Quevedo,
\pl 191 (1987) 282; A. Font, L.E. Ibanez, F. Quevedo, A. Sierra,
\npb 331 (1990) 421
\end{thebibliography}
\end{document}